\documentclass[namedreferences]{SolarPhysics}
\usepackage[optionalrh]{spr-sola-addons} 
\usepackage{graphicx}        
\usepackage{savesym}
\usepackage{stmaryrd}
\usepackage{color}           
\usepackage{url}             
\usepackage{epsfig}

\newcommand{\etal}{{\it et al.}}

\begin{document}

\begin{article}

\begin{opening}

\title{The dependence of the EIT wave velocity on the magnetic field strength}

\author{H.Q.~\surname{Yang}$^{1}$\sep
        P.F.~\surname{Chen}$^{1,2}$
       }
\runningauthor{Yang and Chen}
\runningtitle{Dependence of the EIT wave velocity on the magnetic field strength}

\institute{$^{1}$Department of Astronomy, Nanjing University, Nanjing 210093, China \\
$^{2}$Key Laboratory of Modern Astronomy and Astrophysics (Nanjing University), Ministry of Education, China \\
                     email: \url{chenpf@nju.edu.cn}
       }

\begin{abstract}
``EIT waves" are a wavelike phenomenon propagating in the corona,
which were initially observed in the extreme ultraviolet (EUV)
wavelength by the EUV Imaging Telescope (EIT). Their nature is still
elusive, with the debate between fast-mode wave model and non-wave
model. In order to distinguish between these models, we investigate
the relation between the EIT wave velocity and the local magnetic
field in the corona. It is found that the two parameters show
significant negative correlation in most of the EIT wave fronts,
{\it i.e.}, EIT wave propagates more slowly in the regions
of stronger magnetic field. Such a result poses a big challenge to
the fast-mode wave model, which would predict a strong positive
correlation between the two parameters. However, it is demonstrated
that such a result can be explained by the fieldline stretching
model, \emph{i.e.,} that ``EIT waves" are apparently-propagating
brightenings, which are generated by successive stretching of closed
magnetic field lines pushed by the erupting flux rope during coronal
mass ejections (CMEs).
\end{abstract}


\end{opening}

\section{Introduction}
     \label{S-Introduction}
EIT waves were discovered by the EUV Imaging Telescope (EIT)
\cite{Del95} aboard the Solar and Heliospheric Observatory (SOHO)
initially in the 195 {\AA} channel \cite{moses97,tom99}.
They are transient wavelike disturbances that propagate
almost over the solar disk and are followed by expanding dimmings
(Thompson {\etal}, \citeyear{tom98,tom99}). Later, they were also
identified in other channels like 171 {\AA}, 284 {\AA}, and 304
{\AA} \cite{wills99,zhu04,Dav08}. They originate in around a flare
site and propagate outward, avoiding strong magnetic features and
neutral lines and stopping near coronal holes \cite{tom99} or near
the magnetic separatrix between active regions \cite{Del99}. The
intriguing phenomenon provoked a lot of controversies on the driving
source and its nature ({\it e.g.}, \opencite{Del00}; \opencite{chen08}). It becomes now widely accepted that EIT waves
are physically associated with CMEs, rather than solar flares
\cite{bies02,cliv05,chen06}. Especially, \inlinecite{chen09a} and
\inlinecite{dai10} found that EIT wave fronts are cospatial with CME
frontal loops.

The debate on the nature of EIT waves has been continuing for more
than 10 years. EIT waves were initially explained as the
counterparts of chromospheric Moreton waves \cite{tom99}. Moreton
waves are propagating fronts visible mainly in the H$\alpha$ line
wings, traveling with a velocity of 500-2000 km s$^{-1}$
\cite{Moreton60}. They were successfully explained to be due to
coronal fast-mode waves sweeping the chromosphere ({\it e.g.},
\opencite{Uchida}). Following this line of thought,
\inlinecite{wang00} and \inlinecite{wu01} numerically compared the
propagation of fast-mode waves with the EIT wave observations, and
claimed that the trajectory of the fast-mode wave front matches the
EIT wave very well. The cospatiality of a sharp front in the EIT
image with the H$\alpha$ Moreton wave front in the 1997 September 24
event reinforced such a conjecture \cite{tom00}. However, there
is dispute about the relation between this sharp EIT wave front and the
ensuing diffuse EIT wave fronts. Some authors, {\it e.g.}, 
\inlinecite{warm05}, propose that the sharp front evolves to the diffuse
fronts, whereas others, {\emph e.g.}, \inlinecite{chen05}, suggest that
they are of different origin, with the sharp front being the real
coronal Moreton wave, and the ensuing diffuse fronts being the so-called
EIT wave in the general sense. In their model, the sharp front moved
out of the solar disk and was not visible in the ensuing EIT images
since the cadence of SOHO/EIT, $\sim 15$ min, is too long.

The fast-mode model for EIT waves was first questioned by
\inlinecite{Del99}, who proposed that EIT waves should be related to
magnetic reconfiguration. Based on MHD numerical simulations,
\inlinecite{chen02} proposed that EIT waves are apparently moving
brightenings which are generated by the successive stretching of the
closed field lines, being pushed by the erupting flux rope. The
theory can explain the following observational features: (1) EIT
wave velocity is typically 3 times slower than fast-mode wave
velocity; (2) A substantial outflow is present in the dimming
region, and absent in the EIT wave front \cite{Harra03}; (3) EIT wave
velocity is uncorrelated with type II radio bursts velocity
\cite{Klassen00}. The model is also consistent with the statistical
studies which show that EIT waves are more closely associate with CME
than solar flares \cite{bies02,chen06}. Besides, some authors proposed
alternative models in terms of slow-mode waves \cite{wils07, wang09}
or successive reconnection \cite{attr07}.

According to the fast-wave model, it is expected to see a strong
positive correlation between EIT wave velocity and the local
fast-mode wave velocity or the local magnetic field strength.
However, in the fieldline stretching model of Chen {\etal}
(\citeyear{chen02,chen05}),
the EIT wave velocity is determined by both magnetic field strength
and magnetic configuration \cite{chen09b}, {\it e.g.}, a strongly
stretched configuration as in a quadrupolar field leads to a small EIT
wave velocity. Therefore, it would not see a significant positive
correlation between EIT wave velocity and the local fast-mode wave
velocity or the field strength. In particular, \inlinecite{chen09b}
theoretically showed that as the EIT wave propagates outward just
across the boundary of the source active region, the fast-mode wave
velocity decreases with the distance, whereas the EIT wave velocity
increases. Therefore, it is easy to distinguish the fast-wave model
and the fieldline stretching model by studying the correlation
between EIT wave velocities and the local magnetic field strength,
which is the aim of this paper. Observations and the data analysis
are described in Section 2, the results are presented in Section 3,
and discussions are given in Section 4.

\section{Observations and Data Analysis} 
      \label{S-general}
The two EIT wave events studied in this paper took place on 2007 May
19 and 2007 December 7. Both were located near the solar disk center
in the field of view of the Extreme Ultra Violet Imager (EUVI)
\cite{How08} on board the Solar Terrestrial Relations Observatory
Ahead (STEREO A) satellite, which allows a precise
measurement of the EIT wave velocity.

There are four channels in STEREO/EUVI observations, {\it i.e.}, 171
{\AA} (the formation temperature $T$ $\sim$ 1 MK), 195 {\AA} ($T$
$\sim$ 1.5 MK), 284 {\AA} ($T$ $\sim$ 2 MK), and 304 {\AA}.
It is noted that the 304 {\AA} channel has a coronal
contribution from the Si {\small XI} line ($T$ $\sim$ 1.6 MK), in
addition to the chromospheric line He {\small II} ($T$ $\sim$ 80000 K)
\cite{Bro96,Dav08}. Although EIT waves are more evident in 195 {\AA}
than 171 {\AA} \cite{wills99}, we use the 171 {\AA} images to derive the
EIT wave propagation velocities since the time cadence ($\sim$ 2.5 min)
is much shorter than that of the 195 {\AA} channel ($\sim$ 10 min).

In order to derive the strength of the coronal magnetic field, we
use the potential field extrapolated from the synoptic magnetogram
of Michelson Doppler Imager (MDI) telescope \cite{Sch95} on board the
SOHO satellite with the potential field source surface (PFSS)
technique \cite{SD03}. Since EIT wave fronts are significantly
bright in the low corona ({\it e.g.}, \opencite{cohen09}), the
magnetic field is taken at the height of 0.06$R_\odot$ in the
extrapolated potential field.

\begin{figure}
\begin{center}
\hspace*{0.5cm} \psfig{file=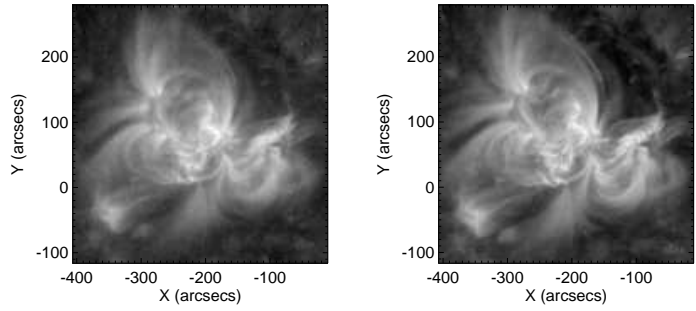,width=9.cm,clip=} \caption{The
image of SOHO/EIT (left) and the corrected image of STEREO/EUVI (right)
at 15:24:00 UT on 2007 May 19 in the bandpass 171 {\AA}. The
correlation coefficient of the two images is 0.96.} \label{fig:519}
\end{center}
\end{figure}

\begin{figure}
\begin{center}
\hspace*{0.5cm} \psfig{file=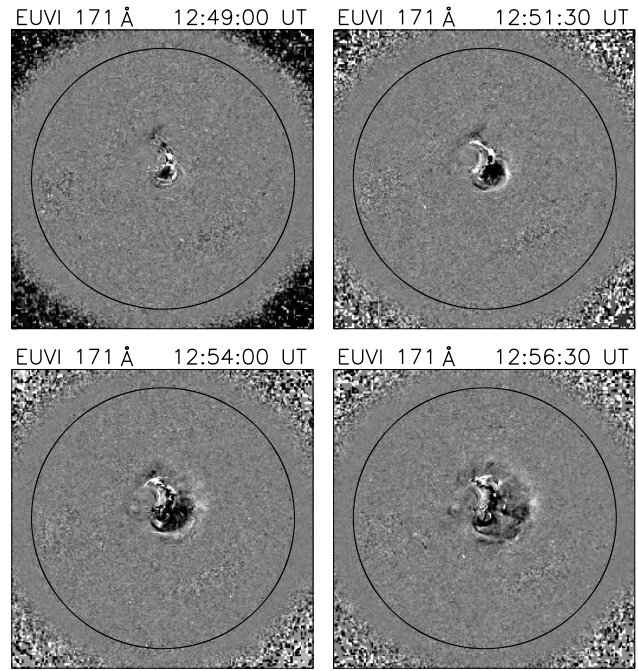,width=9.cm,clip=} \caption{The
base difference images of the EIT wave event at 4 times
from 12:49:00 UT to 12:56:30 UT on 2007 May 19 observed by
STEREO/EUVI showing the wave propagation and the dimming regions on
the solar disk. The black circles mark the solar limb.} \label{diffq}
\end{center}
\end{figure}

\begin{figure}
\begin{center}
\hspace*{0.5cm} \psfig{file=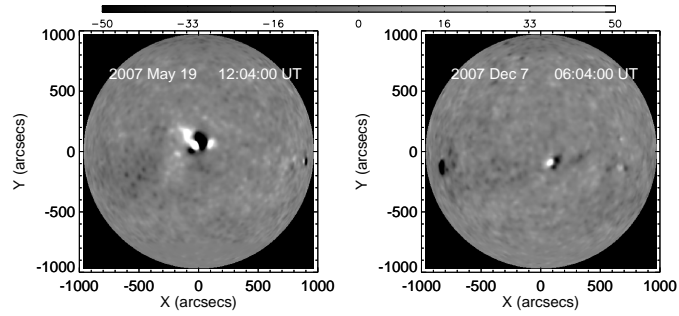,width=9.cm,clip=}
\caption{The
corrected full disk distributions of the radial component of
magnetic field ($B_r$) at the height of $0.06R_\odot$ for the two
EIT events, which are extrapolated from the MDI magnetogram on the
solar surface.} \label{pfssq}
\end{center}
\end{figure}

Since the STEREO/EUVI and SOHO/MDI observe the Sun at different times
and directions, an important step is to coalign these two data sets.
For that purpose, both data sets are coaligned with SOHO/EIT images,
since EUVI and EIT use the same emission lines, while MDI and EIT are
both aboard the SOHO satellite. First, we derotate and remap the images
of EUVI to that of EIT. As an example, the two panels of Figure
\ref{fig:519} display the corrected 171 {\AA} images from
SOHO/EIT (left) and STEREO/EUVI (right) at 15:24:00 UT on 2007 May
18 . The correlation coefficient of the two images is 0.96. After
being remapped to the Earth view, running difference images of
STEREO/EUVI are taken to show the propagation of EIT waves. As an
example, Figure \ref{diffq} depicts the evolution of the EIT wave
event on 2007 May 19. Second, the distribution of the radial
component of the magnetic field ($B_r$) at the height of 0.06
$R_\odot$ is taken from the extrapolated coronal field, and is
derotated and remapped to the same time as the EUVI images. Figure
\ref{pfssq} shows the corresponding $B_r$ distribution for the two
EIT wave events. Note that $B_r$ is the main component of the
magnetic field outside active regions.

\section{Results} 
      \label{S-features}
From the running difference images as shown in Figure \ref{diffq}, we
can manually trace the location of the EIT wave front at each time,
which is illustrated in Figure 4 as solid lines. As the EIT wave
front appears diffuse with time and the magnetic field become weak far
away from the active region, we measure the EIT wave propagation
close to the active region only. The left panel
displays the propagation of the EIT waves at 6 times from 12:46:30
UT to 12:59:00 UT for the 2007 May 19 event, and the right panel
displays the propagation of the EIT waves at 6 times from 04:28:30
UT to 04:41:00 UT for the 2007 December 7 event. Note that only the
evident portion of each EIT wave front is traced, so that the first
event is traced mainly in the west hemisphere, whereas the second
event in the east hemisphere.

\begin{figure}
\begin{center}
\hspace*{0.5cm} \psfig{file=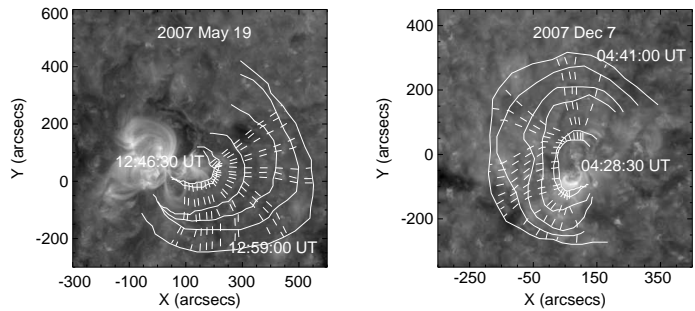,width=9.cm,clip=}
\caption{Huygens plotting of the EIT wave front propagation in the
2007 May 19 event (left) and the 2007 December 7 event (right)
in the bandpass 171 {\AA}. The solid lines correspond to
the wave fronts at different times, while the dashed lines show the
propagating trajectories.} \label{gt}
\end{center}
\end{figure}

In order to derive the propagation velocity, we select 20 starting
points along the first visible EIT wave front, and then track their
trajectories with the Huygens plotting technique as done in
\inlinecite{wills99} and \inlinecite{zhu09}. The trajectories are
displayed as dashed lines in Figure \ref{gt}. The velocity is
obtained by the 3-point central difference scheme. Therefore,
for each trajectory, velocities at 4 times can be obtained. Error in
wave front position is estimated to be $\pm$2 pixels,
following \inlinecite{zhu09}. The local magnetic field at each
position is determined by averaging the surrounding $3 \times 3$
pixels in the extrapolated coronal field at the height of
0.06$R_\odot$ above the solar surface.

\begin{figure}
\begin{center}
\hspace*{0.5cm} \psfig{file=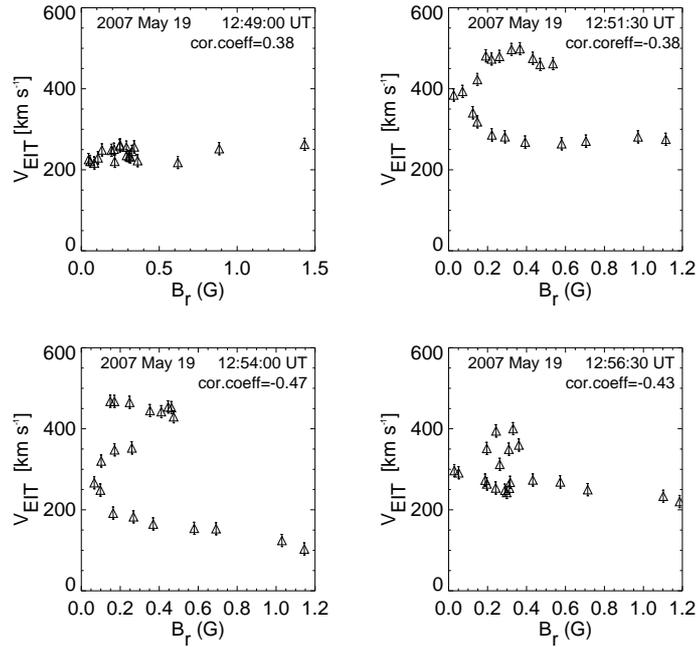,width=9.cm,clip=} \caption{The
relation between EIT wave velocity and magnetic field strength for
the 2007 May 19 event. The four panels correspond to 12:49:00 UT,
12:51:30 UT, 12:54:00 UT, 12:56:30 UT, respectively.}
\label{vcor519}
\end{center}
\end{figure}

The scatter plot of the EIT wave velocity ($v_{\rm EIT}$) vs. the radial
component of the magnetic field ($B_r$) for the 2007 May 19 event is
displayed in Figure \ref{vcor519}, where 4 panels correspond to wave
fronts at 12:49:00 UT, 12:51:30 UT, 12:54:00 UT, and 12:56:30 UT. It
is seen that, only at 12:49:00 UT, a very weak positive correlation
exists between $v_{\rm EIT}$ and $B_r$ (upper left panel), where
$v_{\rm EIT}$ increases slightly as $B_r$ increases by more than 10
times. In the ensuing times, however, $v_{\rm EIT}$ and $B_r$ present a
negative correlation, where EIT waves tend to move more slowly at
the site with a stronger magnetic field.

\begin{figure}
\begin{center}
\hspace*{0.5cm} \psfig{file=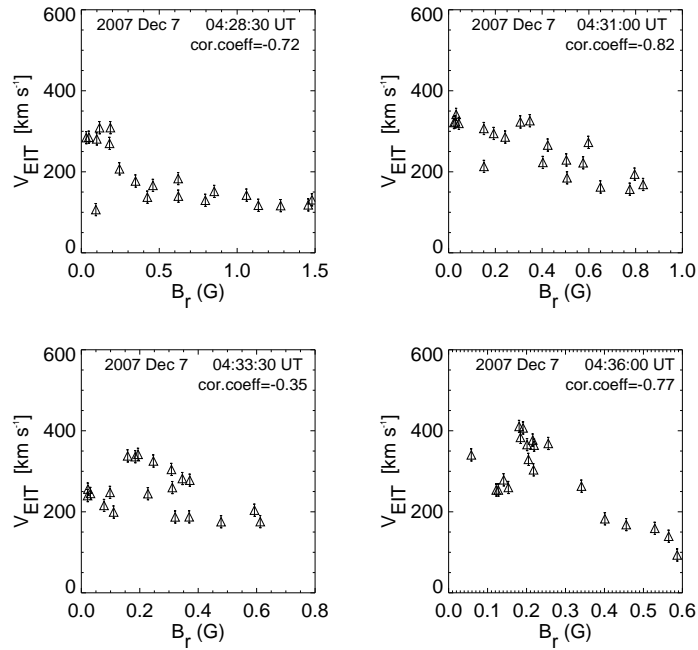,width=9.cm,clip=} \caption{The
relation between the EIT wave velocity and the magnetic field
strength for the 2007 December 7 event. The four panels correspond
to 04:28:30 UT, 04:31:00 UT, 04:33:30 UT, 04:36:00 UT,
respectively.} \label{vcor127}
\end{center}
\end{figure}

The scatter plot of $v_{\rm EIT}$ vs. $B_r$ for the 2007 December 7
event is displayed in Figure \ref{vcor127}, where 4 panels
correspond to the EIT wave fronts at 04:28:30 UT, 04:31:00 UT,
04:33:30 UT, and 04:36:00 UT, respectively. All these panels reveal
that $v_{\rm EIT}$ is negatively correlated to $B_r$, with the
correlation coefficient being smaller than -0.7 at 3 times.

\section{Discussions} 
      \label{S-Conclusion}

``EIT waves" are often explained in terms of fast-mode
magnetoacoustic waves in the corona ({\it e.g.}, \opencite{wang00};
\opencite{wu01}; \opencite{vrs02}; \opencite{war04};
\opencite{gre08}; \opencite{pom08}; \opencite{gop09};
\opencite{pat09}). However, the wave model cannot explain the
following features of ``EIT waves" (see \opencite{wils07};
\opencite{chen08}, for details): (1) the ``EIT wave" velocity is
significantly smaller than those of Moreton waves. The latter are
well established to be due to fast-mode waves in the corona;
(2) the ``EIT wave" velocities have no correlation with
those of type II radio bursts \cite{Klassen00}; (3) the ``EIT wave"
fronts may stop when they meet with magnetic separatrices
\cite{Del99}; (4) the ¡°EIT wave¡± velocity may be below 100 km
s$^{-1}$ ({\it e.g.}, \opencite{Dav08}), which is even smaller than
the sound speed in the corona. Moreover, the rotation of EIT wave
fronts reported by \inlinecite{pod05} was found to be linked to the
filament rotation, implying that EIT waves might not be fast-mode
waves \cite{attr07}.
Note that it may be argued that the lack of correlation between
the velocities of EIT waves and type II radio bursts is due to that the
former is located at the leg, while the latter is at the top of the
same fast-mode wave. However, the velocities of Moreton waves and
type II radio bursts do show linear correlation \cite{pint77}, although
Moreton wave is also located at the leg of the type II radio 
burst-related shock wave.

In order to distinguish whether EIT waves are fast-mode waves or
not, we investigated the relation between the EIT wave velocity and
the local magnetic field with the high-cadence observations from
STEREO/EUVI. With the Huygens plotting technique, EIT wave velocity
distribution along each front at 4 times from two EIT wave events
was derived, it is found that except at one moment when the EIT wave
velocity ($v_{\rm EIT}$) is slightly positively correlated with the
local magnetic field ($B_r$), $v_{\rm EIT}$ at all other moments is
negatively correlated with $B_r$, and at some moments, the negative
correlation is rather significant. In order to examine the validity
of the result, we also calculated the relation between $v_{\rm EIT}$ and
$B_r$, with $B_r$ being taken at heights like 0.3$R_\odot$ and
0.1$R_\odot$, it is found that the result does not change. Such a
result poses a big challenge to the fast-mode wave model for EIT
waves, since the model would predict a strong positive correlation
between $v_{\rm EIT}$ and $B_r$.

\begin{figure}
\begin{center}
\hspace*{0.5cm}
\psfig{file=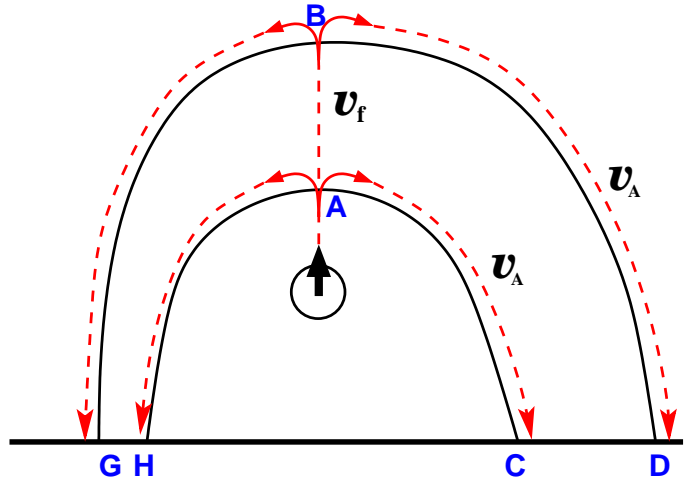,width=9.cm,clip=}
\caption{Sketch of the fieldline stretching model for EIT waves
proposed by Chen et al. (2002, 2005), where solid lines are the
magnetic field, and dashed lines indicate how the magnetic fieldline
stretching is transferred with the local fast-mode wave speed, {\it 
i.e.}, the Alfv\'en speed $v_A$ along the field line and $v_f=
\sqrt{v_A^2+v_s^2}$ perpendicular to the field line, where $v_s$ is the
sound speed.}
\label{cartoon}
\end{center}
\end{figure}

It would be then interesting to check whether the significant
negative correlation between $v_{\rm EIT}$ and $B_r$ can be explained by
the fieldline stretching model proposed by Chen {\etal}
(\citeyear{chen02,chen05}). The fieldline stretching model is
illustrated in Figure \ref{cartoon}: as the flux rope is ejected
upward, all the overlying field lines will be pushed to stretch up,
and for each fieldline, the stretching starts from the top part. That
is to say, the perturbation propagates from point A to C with the
fast-mode wave speed, by which the EIT wave front reaches point C. Note
that the fast-mode wave speed along the magnetic field line is equal to
the Alfv\'en speed $v_A$. Then the perturbation propagates from point A
to B and then to D with the local fast-mode wave speed $v_f=\sqrt{v_A^2
+v_s^2}$ (here $v_s$ is the sound speed), by which a second front
reaches point D. The apparent velocity of the EIT wave is the distance
CD divided by the time difference of the perturbation transfer, {\it
i.e.,}

\begin{equation}
v_{\rm EIT}=CD/\Delta t,\label{equ}
\end{equation}

\noindent
where $\Delta t=\int_A^B1/v_f ds+\int_B^D1/v_A ds-\int_A^C1/v_A ds$.
According to this model, we can calculate the EIT
wave velocity profile for any magnetic configuration.

\begin{figure}
\begin{center}
\hspace*{0.5cm}
\psfig{file=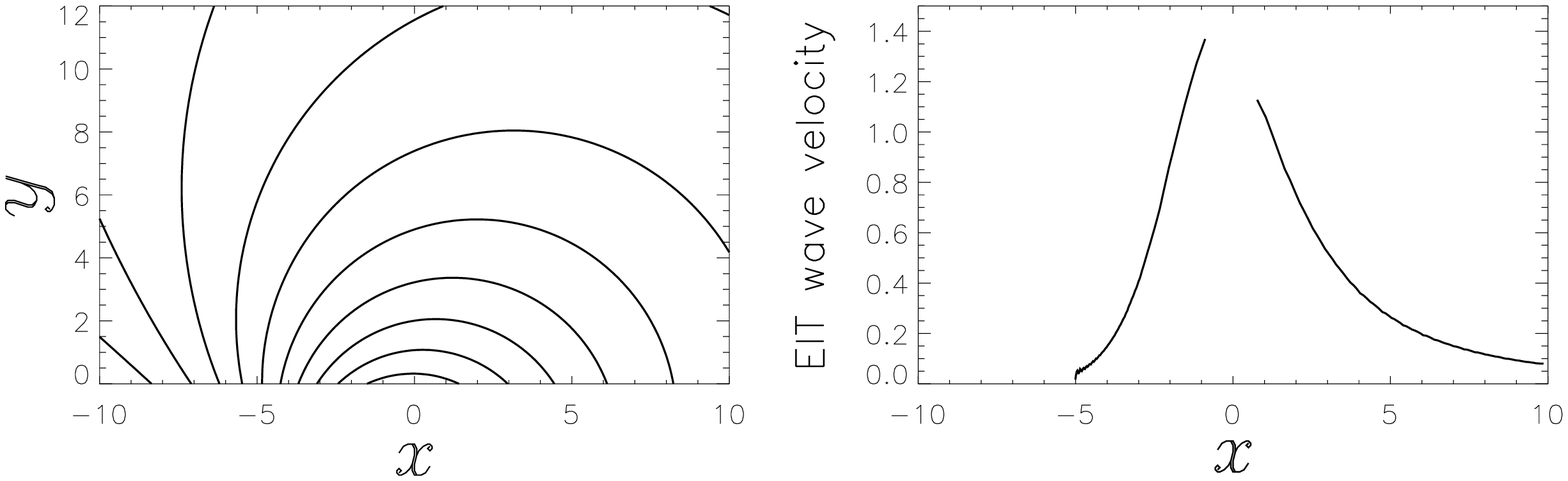,width=10.cm}
\caption{Left: an asymmetric magnetic configuration corresponding to
an oblique magnetic dipole below the solar surface; Right: the
corresponding EIT wave velocity distribution along the horizontal
distance. The velocity is in unit of 900 km s$^{-1}$.}
 \label{num}
\end{center}
\end{figure}

In order to get an asymmetric magnetic configuration, we put an
oblique magnetic dipole below the solar surface, with a line current
(along the $z$-direction) at $x=-2.4$  and $y=-5.5$ , and another
line current (along the negative $z$-direction) at $x=-2.6$  and
$y=-5.7$. The magnetic field is plotted in the left panel of Figure
\ref{num}, with the typical magnetic field being $\sim 3.3$ G.
From the figure we can see that the
magnetic field left to the magnetic neutral line is stronger than
that to the right.  For simplicity, we assume a uniform and
isothermal plasma with the plasma number density of $10^8$ cm$^{-3}$
and temperature of $10^6$ MK. The resulting typical Alfv\'en speed is
$v_{A0}$=900 km s$^{-1}$. The corresponding EIT wave velocity profile,
which is calculated with Equation (\ref{equ}), is depicted in the right
panel of Figure \ref{num}, where the wave velocity is in unit of
$v_{A0}$. It is seen that the EIT wave velocity profile
is complicated. Within $|x|\leq$ 2, the EIT wave velocity
($v_{\rm EIT}$) is higher on the left side, whereas as $|x|$ is larger
than 2, $v_{\rm EIT}$ becomes higher on the right side. This means that
for the magnetic field in the left panel of Figure \ref{num},
$v_{\rm EIT}$ and $B_r$ are correlated positively near the neutral line
and negatively further out. Such a result is qualitatively
consistent with the observed features presented in Section 3.

By comparing the top two panels in either Figure \ref{vcor519} or Figure
\ref{vcor127}, it can also be seen that both events analyzed in this
paper show that EIT wave velocity increases as the front propagates
outward near the boundary of the source active region. This confirms
the result presented in \inlinecite{Dav08}. \inlinecite{chen09b}
claimed that this feature poses another big challenge for the fast-mode
wave model for EIT waves, which implies a deceleration of EIT wave
velocity away from the active region since the fast-mode wave speed is
much larger near the active region than in the quiet region.
\inlinecite{chen09b} also demonstrated that the acceleration can be well
accounted for in the fieldline stretching model. The reason is quite
straightforward: according to their model
\cite{chen02,chen05}, as illustrated in Figure \ref{cartoon}, if the
magnetic configuration is strongly stretched, {\it i.e.,} the distance
between points B and C is larger, the apparent EIT wave velocity derived
by Equation (\ref{equ}) would be smaller. The boundary of active regions
does present such a kind of stretched magnetic configuration, as
demonstrated in \inlinecite{chen09b}. \inlinecite{vero08} also analyzed
the 2007 May 19 event, however, they claimed that the EIT wave velocity
decreases with the distance from the source active region as expected
from the fast-mode wave model. The discrepancy between their results
and ours, as well as in \inlinecite{Dav08}, is probably due to that they
did not show the EIT wave propagation before 12:52:00 UT, when the EIT
wave velocity was increasing.

Note that we focused our study on the EIT wave propagation near the
source active region, where the significant variation of the EIT wave
velocity happens. In the quiet region far from the source active region,
the EIT wave velocity becomes relatively stable \cite{ma09}. With a
suitable choice of the coronal density model, it would be easier to
reproduce the EIT wave propagation in the large scale as done by
\inlinecite{wang00} and \inlinecite{wu01}, especially when there were
only $\sim3$ snapshots of EIT wave propagation in the SOHO era. However,
it is hard to imagine that the negative correlation between EIT wave
velocity and the local magnetic field strength near the boundary of the
source active region can be reproduced in the fast-mode wave framework. 

To summarize, we analyzed the relation between EIT wave velocity
($v_{\rm EIT}$) and the local magnetic field ($B_r$) in two events
observed by STEREO/EUVI with a high cadence. It is found that
$v_{\rm EIT}$ and $B_r$ are negatively correlated in most of the fronts,
except along one front where $v_{\rm EIT}$ and $B_r$ show a weak
positive correlation. It is further revealed that such features,
which pose a big challenge for the fast-wave model for EIT waves,
can be well explained by the fieldline stretching model proposed by
Chen {\etal} (\citeyear{chen02,chen05}).

\begin{acks}
The research is supported by the Chinese foundations 2006CB806302 and
NSFC (10403003, 10933003, and 10673004). We thank the referee for
constructive comments that helped to improve the paper and M. L. Derosa
for the help on the use of PFSS routine. The SECCHI data used here were
produced by an international consortium of USA, UK, Germany, Belgium,
and France. {\it SOHO} is a project of international cooperation between
ESA and NASA.
\end{acks}

\bibliography{SOLA_bibliography_example}

\end{article}

\end{document}